# Magnetodielectric consequences of phase separation in the colossal magnetoresistance manganite $Pr_{0.7}Ca_{0.3}MnO_3$

R. S. Freitas[1], J. F. Mitchell[2], and P. Schiffer[1*]

[1]Department of Physics and Materials Research Institute, Pennsylvania State University, University Park, Pennsylvania 16802

[2]Materials Science Division, Argonne National Laboratory, Argonne, Illinois 60439

**ABSTRACT**

We have studied the low-frequency dielectric properties of the phase-separated manganite $Pr_{0.7}Ca_{0.3}MnO_3$ as a function of applied magnetic field in the low temperature phase-separated state. The dielectric constant is strongly field dependent and also depends on the magnetic field history of the sample. The dielectric behavior appears to be associated with the hopping of polaronic charge carriers, and we can derive the field dependent hopping energy barrier from the frequency dependence of the dielectric constant. This analysis allows us to associate the metal-insulator transition observed in this material with the field-induced suppression of the polaron activation energy.

[*] Corresponding author: schiffer@phys.psu.edu

I. INTRODUCTION

The study of materials where strong correlations among electrons plays a crucial role defining the macroscopic properties has attracted a great deal of interest since the discovery of the high temperature superconductivity in cuprates[1] and the colossal magnetoresistance in manganites.[2] One of the most interesting properties of these materials is the occurrence of spontaneous phase separation between different electronic and magnetic states. This sort of phase-separated (PS) state is a common characteristic of many manganites, resulting from the strong coupling between lattice, electronic and magnetic degrees of freedom.[3] From the magnetic point of view, the PS state in the manganites often results in glassy behavior, and several studies have reported unusual relaxation dynamics and frequency dependent phenomena.[4,5,6] While the magnetic glassiness associated with phase separation has drawn considerable attention, there has been less focus on the possibility of electronic glassiness associated with phase separation and how the frequency dependent dielectric properties of the manganites couple with the magnetic properties in the phase-separated state. This subject is especially interesting in light of a suggested electronic glassy state on the dielectric response of hole-doped cuprates and nickelates.[7]

One manganite material of particular interest is $Pr_{0.7}Ca_{0.3}MnO_3$, since it has been shown to exhibit a variety of unusual behavior associated with the phase-separated state. This low-bandwidth compound is insulating at zero magnetic field, and has a charge ordering transition around 225 K followed by antiferromagnetic and ferromagnetic transitions at 130 and 110 K.[8] Interestingly, a metallic phase can be induced by the application of magnetic fields,[9] light,[10] pressure,[11] high electric field,[12] or even irradiation by x-rays.[13] Recent neutron scattering studies have established that the low temperature ground state of this system is microscopically inhomogeneous with the coexistence of ferromagnetic (FM) and charge-ordered antiferromagnetic (CO-AF) regions.[14] The application of a magnetic field first drives the system into a phase-separated insulating state with above 80% of the full saturated ferromagnetic state at 1 T and then to a ferromagnetic metallic state above 4 T with the full saturated moment.[4,9] Glassiness has been observed in the magnetic susceptibility of this material in the form of frequency-dependent ac magnetic susceptibility and differences between the field-cooled and zero-field-cooled magnetization, although the existence of a conventional spin-glass phase at low temperatures was excluded.[4]

We have studied the coupling of the magnetic and dielectric properties of single crystalline $Pr_{0.7}Ca_{0.3}MnO_3$ at low temperatures in the phase-separated state by measuring the dielectric susceptibility as a function of temperature, magnetic field, and frequency. We find that the dielectric susceptibility displays a difference between the field-cooled and zero-field cooled state, in analogy to the magnetization. Dielectric relaxation associated with localized hopping of polaronic charge carriers leads to a characteristic frequency dependence, which we analyze to determine the characteristic energy barrier to polaron hopping. By tracking the evolution of this frequency-dependent behavior with magnetic field, we are able to elucidate the field-dependence of polaron relaxation.

## II. EXPERIMENT

We studied high-quality single-crystalline samples of $Pr_{0.7}Ca_{0.3}MnO_3$ grown in a floating zone image furnace. Additional information about the magnetic and thermodynamic properties of these samples can be found on Refs. 4 and 15, respectively. Dielectric measurements were made with an ultra-precision capacitance bridge (AH 2700A) for frequencies varying from 50 Hz to 20 kHz. The temperature and field dependent dielectric measurements were made using a commercial cryostat (Quantum Design PPMS). The sample was additionally characterized by transport measurements, made by a conventional dc four-probe method, and by magnetometry using a superconducting quantum interference device (SQUID) magnetometer (Quantum Design MPMS). For the dielectric measurements, we measured 140-μm-thick disks with circular electrodes of diameter of 0.55 mm evaporated on the surfaces. To test the influence of the contact on the capacitance we used different configurations, from evaporated gold with or without preevaporated chromium or titanium as adhesive layers, to contacts made directly with silver paint. We also conducted control experiments with electrodes isolated from the sample with a 10 μm thick layer of $Al_2O_3$.

III. RESULTS AND DISCUSSION

Figure 1 shows the measured temperature dependence of the dielectric constant (ε') for different contact types at a typical low frequency of $f$ = 100 Hz. The qualitative behavior of the dielectric constant is similar for all of the different contact materials, showing a weak decrease with decreasing temperature down to $T$ = 60 K, and a step like drop around 40 K. The low temperature value of the dielectric constant, $\varepsilon(T\rightarrow 0)$ = 57, is independent of the contact type, temperature and frequency (see Fig. 3(a)), reflecting an intrinsic property of the material, and this value is in agreement with the reported for other manganites.[16] Above 40 K the dielectric constant is highly dependent on the contact nature assuming values as high as $\varepsilon'$ = $10^4$. It has been proposed[17] that these high values may not be intrinsic, but probably related to the depletion layers, i.e., Schottky barriers formed in the region of the sample close to the metal electrode. The inset of Fig 1 shows the results for the case where the electrode is isolated from the sample by a thin layer of $Al_2O_3$. For this configuration, no Schottky barriers are formed and the dielectric constant values are much smaller than before. As can be clearly seen on the figure, the step-like decrease is still present at the same temperature as observed before, which indicates that this feature represents an intrinsic attribute of the sample. In the remainder of the paper only results for the sample with evaporated Ti/Au contacts will be shown. We emphasize that similar results were obtained for all contact configurations, and in particular the different contacts result in a < 10% change in the derived polaron energy and its field dependence (discussed below).

Figure 2 displays the field-cooled (FC) and zero-field-cooled (ZFC) temperature dependence of the dielectric constant (Fig. 2(a)) and magnetization (Fig. 2(b)). Both quantities show history dependence with separation between FC and ZFC at low temperatures. For the dielectric data, the irreversibility temperatures at which the ZFC and FC curves merge are somewhat below those for the magnetization data, but they both decrease with increasing field. The inset shows the field dependence of the relative differences $\Delta M = (M_{FC}-M_{ZFC})/M_{FC}$ and $\Delta\varepsilon = (\varepsilon'_{FC}-\varepsilon'_{ZFC})/\varepsilon'_{FC}$ at $T$ = 4 K. At low fields, H < 0.5 T, $\Delta M$ is large and strongly suppressed with increasing field while $\Delta\varepsilon$ increases steadily. The behavior of $\Delta M$ in this range of fields is attributable to changes in the ferromagnetic clusters with the application of the field, and it is in this regime that the magnetization increases sharply to a large fraction of the total moment.[4] The fact that $\Delta M$ and $\Delta \varepsilon$ do not seem to be correlated in this regime suggests that $\varepsilon$ is not controlled by the ferromagnetic portions of the sample. Above 0.5 T the dielectric response is much more affected by the magnetic field history of the sample than the magnetization ($\Delta \varepsilon >> \Delta M$). The qualitative field dependence of $\Delta M$ and $\Delta \varepsilon$ are quite similar; both increase with magnetic field up to a maximum value around $H = 2.5$ T and diminish for higher fields. The origins of this behavior are found in field-induced changes in the energy barrier to polaron hopping, discussed in detail below.

We now consider the step-like anomaly in the temperature dependence of the dielectric constant. Figure 3(a) shows the temperature dependence of the dielectric constant and dielectric loss (Fig. 3(a), inset) measured at three different frequencies. The sharp drop of the capacitance coincides with a peak in the dielectric loss, which shifts to higher temperatures with increasing frequency. When plotted as a function of frequency at constant temperature, the dielectric loss thus displays a peak that shifts to higher frequencies with increasing temperature (see Fig. 4). The presence of such a peak indicates that there is a characteristic charge relaxation process with a relaxation time that corresponds to the inverse of the peak frequency. Similar behavior has been reported for other manganites[16,18,19] and different perovskite systems,[20,21,22] and it has been attributed to localized hopping of polarons between lattice sites with a characteristic timescale. Figure 3(b) shows the frequency dependence of the dielectric loss measured under different applied magnetic fields at 30 K. Here the dielectric loss is scaled to its maximum value as a function of the scaled frequency at the same point, and a solid line shows the expected Debye behavior if there were a single relaxation time for the system.[23] The slight broadening of the peak relative to this form indicates that there is a narrow distribution of relaxation times, as will be discussed below. The application of magnetic fields using the field-cooling mode broadens the loss peak, suggesting an increase in the distribution of relaxation times discussed below.

In order to understand the underlying nature of the relaxation process, we have measured the frequency dependence of the dielectric loss at different temperatures and magnetic fields. Figure 4 demonstrates that the data can be well described using the Cole-Cole expression: [24]

$$\varepsilon = \varepsilon_\infty + \frac{\varepsilon_0 - \varepsilon_\infty}{1 + (i\omega\tau)^{1-\beta}} \tag{1}$$

where $\varepsilon = \varepsilon' + i\varepsilon''$ is the complex dielectric constant, $\varepsilon_0$ and $\varepsilon_\infty$ are the values of the dielectric constant in the low- and high-frequency limit, $\omega$ is the angular frequency, $\tau$ is the characteristic relaxation time and the phenomenological parameter $\beta$ is a measurement of the relaxation broadening. For $\beta = 0$, this expression reduces to the Debye single relaxation process. In the fits shown in Fig. 4, this parameter increases with magnetic field from 0.31 to 0.50 between 0 and 2.4 tesla at 30 K. This increase in the value of $\beta$ could suggest the development of correlations among the relaxation units with the application of field,[25,26] or simply an increase in the distribution of energy barriers which would not be surprising in this electronically disordered system.

The relaxation times resulting from the fitting of $\varepsilon''(\omega)$ display thermally activated behavior which can be fit to an Arrhenius law $\tau = \tau_0 \exp[E_a/(k_B T)]$ (Fig 5, lower inset), where $E_a$ is the activation energy (corresponding to the barrier to polaron hopping), $k_B$ is the Boltzmann constant, and $\tau_0$ is an attempt time, which increases from 0.52 to 11.3 x $10^{-8}$ s from 0 to 3 tesla. These values are in agreement with those reported for other manganites.[16] At zero magnetic field, $E_a/k_B = 377$ K, which is of the same order as the charge order temperature $T_{co} \sim 200 - 250$ K,[8] supporting the polaronic charge carrier origin of the relaxations. Our dc resistivity measurements are also consistent with polaronic hopping at low temperatures. On the upper inset of Fig. 5 we show the resistivity data in the temperature range 45 - 120 K, scaled to the Mott's variable-range-hopping (VRH) model $\ln(\rho/\rho_0) = (T_0/T)^{1/4}$.[27] In the VRH mechanism the hopping energy is given by: $E_{Mott} = 0.25 k_B T_0^{1/4} T^{3/4}$, and using the $T_0$ value from the VRH fitting of the resistivity measured at zero magnetic field and $T = 35$ K, an average temperature around which the dielectric relaxation was measured, we obtain from Eq. (2) $E_{mott}/k_B = 389$ K. This value is in excellent agreement with the activation energy estimated above from the ac dielectric relaxation, although it should be noted that there is some arbitrariness in our choice of temperature, and the range of temperatures from 20 - 40 K would give values of $E_{Mott}/k_B = 255 - 430$ K. This range of values is in agreement with the activation energy obtained by the dielectric relaxation, providing further evidence that the steplike drop of the dielectric constant is an intrinsic attribute of the sample. We have also tried to fit the resistivity with both a band-gap model and a model of nearest-neighbor hopping of small polarons (both of which have been used for the high-temperature phase of manganites),[28] but we find that the data best follow the $T^{-1/4}$ law, supporting the VRH model as the mechanism for current transport at low temperatures.

As mentioned above, the application of a magnetic field to zero-field-cooled $Pr_{0.7}Ca_{0.3}MnO_3$ converts the magnetoelectronic behavior to predominantly ferromagnetic and insulating above ~ 0.5 tesla and then to a ferromagnetic metallic state above ~ 4 tesla. We now consider the effects of applied magnetic field on the dielectric properties in more detail, with the goal of gaining a better understanding of how the magnetoelectronic state evolves with applied magnetic field at low temperature. As shown in Fig. 5, the polaron activation energy obtained from the dielectric relaxation under the FC procedure ($\varepsilon_{FC}$ in Fig. 5) decreases linearly with magnetic fields and extrapolates to zero at $H \approx 5$ T, which is close to the field required for inducing the first order transition to a metallic state at low temperatures.[29] The activation energy deduced from the dielectric relaxation under ZFC process ($\varepsilon_{ZFC}$ in Fig. 5) is approximately independent of field below $H \cong 0.5$ T, and then decreases with increasing field, attaining a slope similar to that of the FC data for $H > 1$ T. Since the activation energy for the polaronic hopping is higher for the ZFC case, the respective dielectric response should be smaller than the FC one, which is exactly the behavior we observed (see Fig. 2(a)). We can therefore understand the FC/ZFC difference in the dielectric constant as a direct consequence of the decrease of polaron activation energy with the application of magnetic fields.

The validity of extracting a polaron activation energy from our data is supported by both the resistivity data and dielectric measurements taken on samples with difference contacts. The activation energy extracted from the dc resistivity fits is obtained at temperatures above the bifurcation between FC and ZFC magnetization, and thus it is not affected by the field cooling protocol. The activation energy is very near to that from the dielectric data, strongly supporting the polaronic interpretation of the data. Further evidence for a polaronic interpretation comes in Figure 6, which shows the field dependence of the activation energy obtained under FC process for samples with different contacts. As can be clearly seen on the figure, the value of the activation energy is virtually the same for all different contact configurations as one would expect if the behavior were due to the sample properties rather than some artifact of the contacts. Furthermore, the field dependence of the activation energy does not depend on the nature of the electrical contact. This gives strong evidence that the dielectric relaxation observed here is in fact an intrinsic property of the sample and not contact related.

We now consider the functional form of $E_a(H)$. The magnetic state of the sample changes drastically from zero field to 1 tesla, with the moment rising to ~80% of the full ferromagnetic saturation. This rise in the magnetization is associated with a large portion of the sample being in a ferromagnetic insulating phase with the remainder persisting in a charge-ordered antiferromagnetic state.[4,14] The almost constant activation energy found in this range of magnetic fields indicates that the dielectric response is essentially independent of the magnetization of the sample as a whole (note that even the FC $E_a(H)$ data change linearly, and by only around 20% in this field range), and the difference between the behavior of $E_a(H)$ for the FC and ZFC cases below 1T is possibly due to differences in the morphology of the phase separation in the two different states.

The linear dependence of $E_a(H)$ at higher magnetic fields is observed for both field-cooled and zero-field-cooled data and could be associated with double-exchange allowing for easier hopping as the sample becomes more magnetized or a changes in the fractions of the sample in the different phases. The fact that the dependence is linear with field adds further evidence that the dielectric properties are dominated by that portion of the sample which is not in the ferromagnetic state (where the moment nearly saturates at H ~ 1 T). The linear suppression of $E_a(H)$ extrapolates to zero at H~5 tesla. This is very interesting, since it implies that the first order insulator-metal transition which is observed near 4 tesla can be attributed to a suppression of the polaron hopping energy barrier. The nature of this transition (which is from a nearly ferromagnetic state to a fully ferromagnetic state) has been the subject of considerable interest,[30] and our data indicate it to be associated with field-induced effects on correlated electronic behavior in the charge-ordered antiferromagnetic portion of the sample rather than through percolation of a ferromagnetic metallic phase (as has been convincingly demonstrated in other phase-separated materials[31,32]). The non-percolative nature of the field-induced phase transition is consistent with previous studies on the $Pr_{0.7}Ca_{0.3}MnO_3$ system that demonstrate the transition is strongly first order.[15,30] It is worth noting that these studies, particularly that of Fernandez-Baca *et al.*,[30] explicitly monitor only the ferromagnetic portion of the sample. Our observations of changes in the charge ordered insulator complement those of Fernandez-Baca *et al.* and underscore the close coupling between the different magnetoelectronic phases.

IV. CONCLUSIONS

We have investigated the low temperature insulating state of the phase-separated manganite $Pr_{0.7}Ca_{0.3}MnO_3$. The dielectric properties are strongly affected by the application of a magnetic field, and the history of the field application. The correlation between the activation energy derived from the dielectric measurements and the temperature dependence of the resistivity indicates that the variable range hopping dominates both ac and dc electric transport on this material, and there is no evidence for the broad range of relaxation times which might be associated with electronic glassiness. The magnetic field dependence of the polaron activation energy suggests that the polaron hopping is relatively insensitive to the total magnetization of the sample, and is dominated by that portion of the sample which is not ferromagnetically ordered at low fields. The field dependence also suggests that the first order metal-insulator transition results from the field-induced suppression of the polaron activation energy. This finding indicates that the field-induced effects on phase-separated states in manganites are not simply in changing the relative fractions of the different phases, but also in subtle alterations of the properties of each of the two phases.

ACKNOWLEDGMENTS

We gratefully acknowledge useful discussions with D. N. Argyriou and sample preparation assistance from Hong Zheng and John Pearson. P. S. and R.S.F. acknowledge support from NSF grant DMR-0401486 and DMR-0213623. J.F.M. acknowledges support from the U.S. Department of Energy, Office of Science under Contract No. W-31-109-ENG-38. R.S.F. thanks the CNPq-Brazil for sponsorship.

**Figure captions:**

FIG. 1. (Color online) Temperature dependence of the dielectric constant measured at 100 Hz for different contact types. The inset shows the data measured with insulating $Al_2O_3$ layers between the electrode and the sample.

FIG. 2. (Color online) Field cooled (open symbols) and zero field cooled (closed symbols) temperature dependence of (a) dielectric constant and (b) magnetization. For $H$ = 1.2 T and 1.8 T the dielectric data are shifted for clarity. The inset shows the field dependence of the difference $\Delta\varepsilon = (\varepsilon'_{FC}-\varepsilon'_{ZFC})/\varepsilon'_{FC}$ and $\Delta M = (M_{FC}-M_{ZFC})/M_{FC}$ at 4 K.

FIG. 3. (Color online) (a) Temperature dependence of the dielectric constant ε' and dielectric loss ε" measured with different frequencies. (b) Dielectric loss scaled to its maximum value as a function of the scaled frequency for different magnetic fields at 30 K. The solid line represents the Debye behavior expected for a single relaxation time.

FIG. 4. (Color online) Frequency dependence of the dielectric loss for selected temperatures and magnetic fields illustrating the dielectric relaxation. The lines are fits to the Cole-Cole function as described in the text.

FIG. 5. (Color online) Field dependence of the activation energy $E_a$ obtained from ac dielectric relaxation under field cooling ($\varepsilon_{FC}$) and zero-filed cooling ($\varepsilon_{FC}$) procedures; and from the dc resistivity under field cooling ($\rho_{dc}$). The upper and lower inset show the dc resistivity scaled to the VRH model and the Arrhenius behavior of the relaxation times at different magnetic fields. The dotted lines are guide to the eyes.

FIG. 6 (Color online) Field dependence of the activation energy $E_a$ obtained from ac dielectric relaxation under field cooling procedure for samples with different contact types.

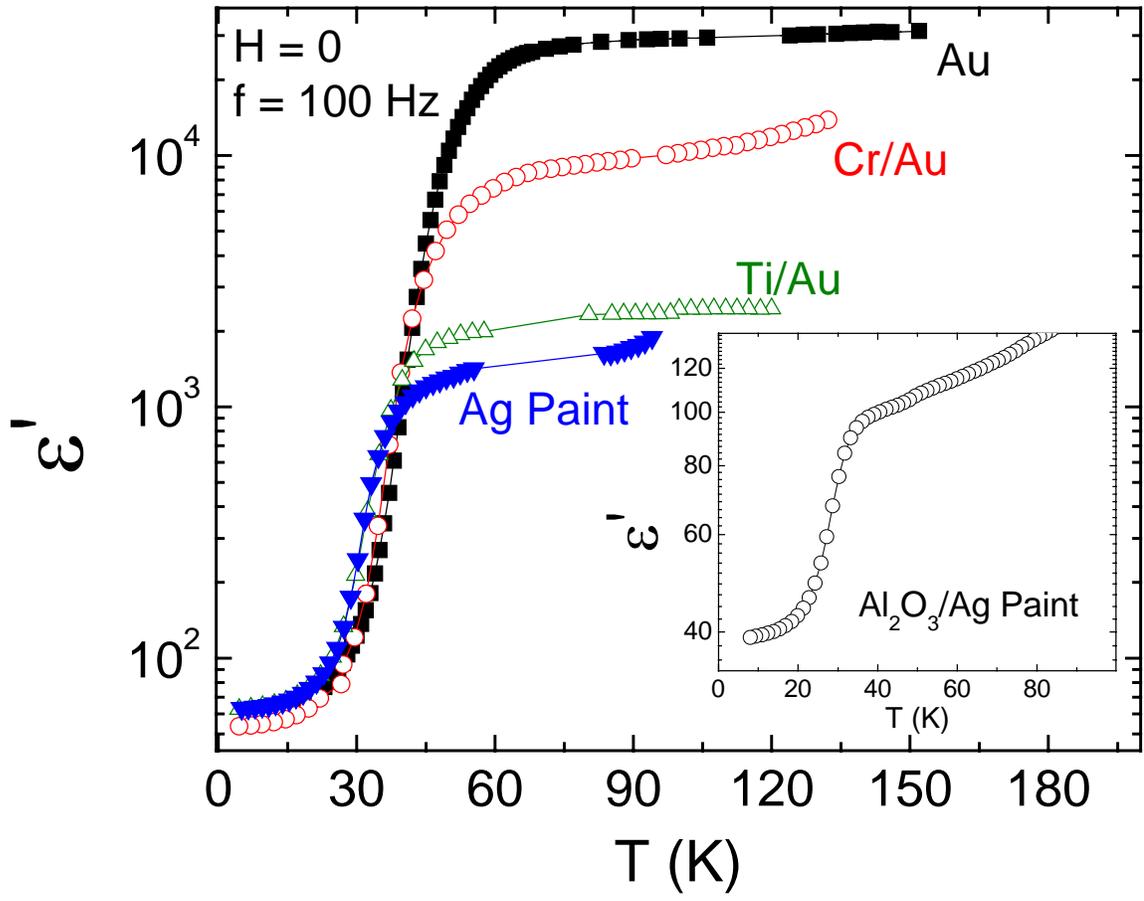

Fig.1 Freitas et al.

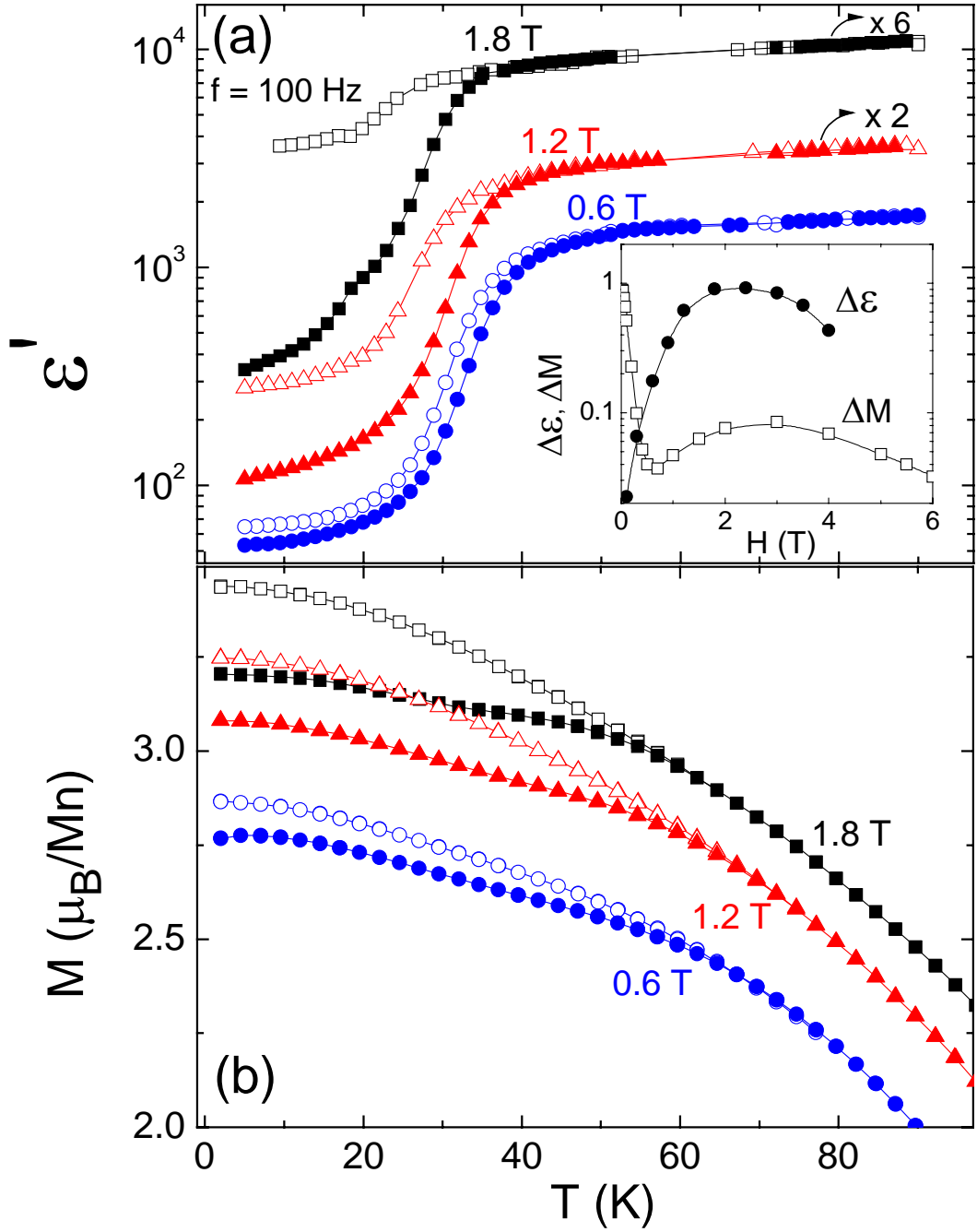

Fig.2 Freitas et al.

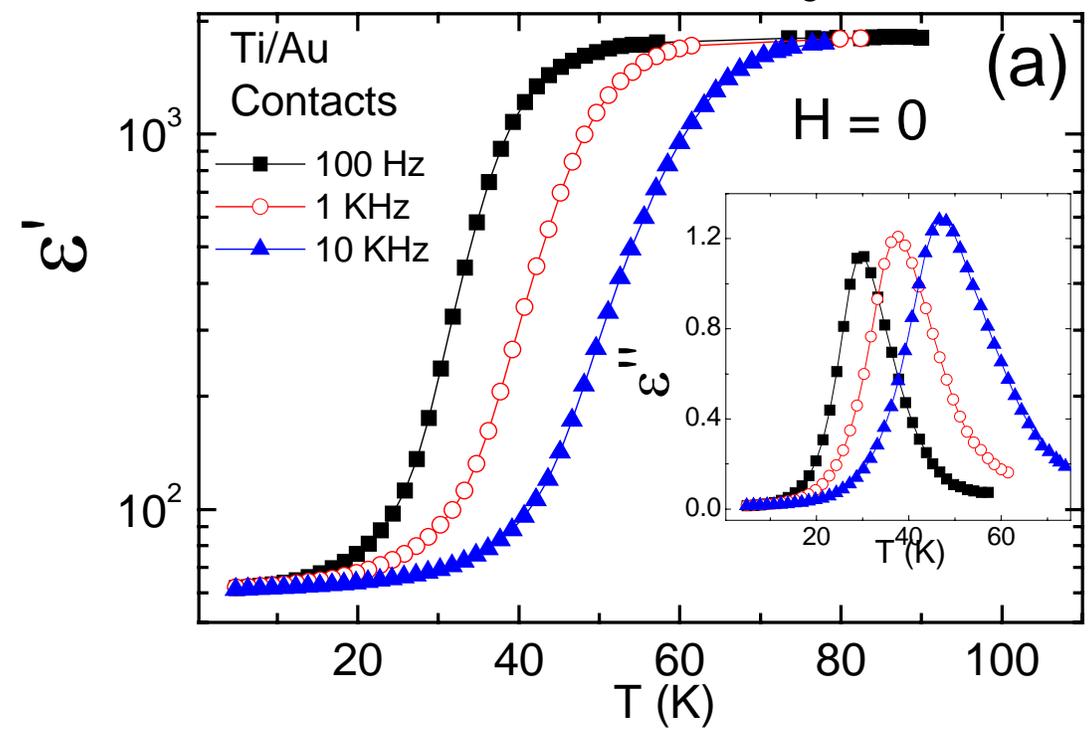
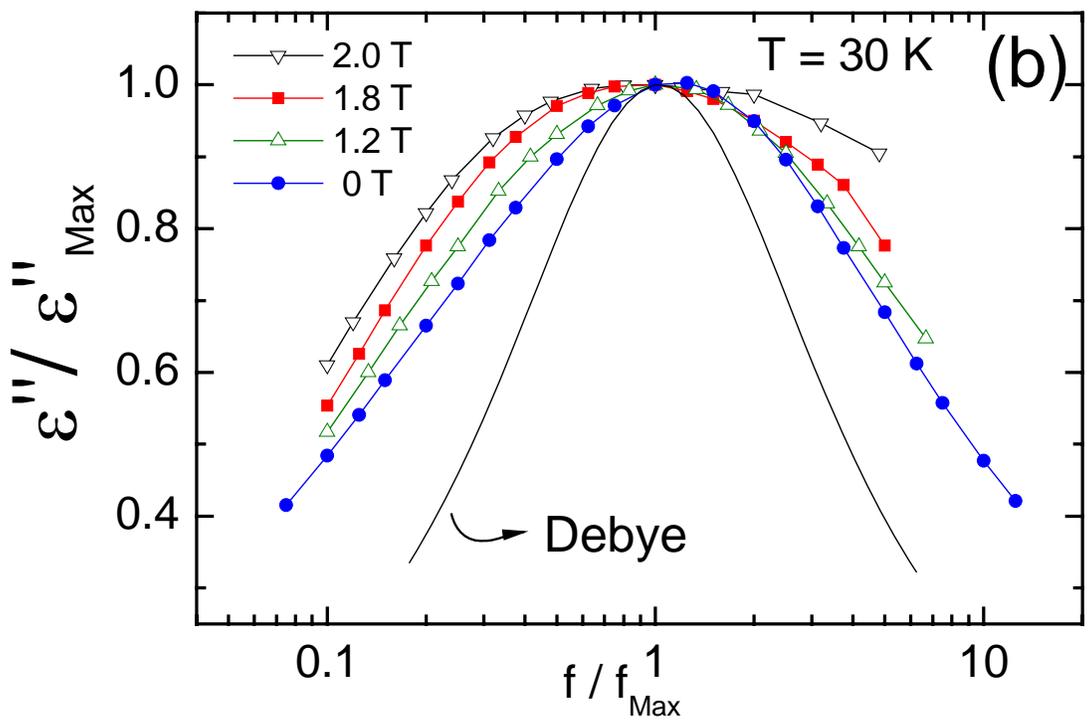

Fig.3 Freitas et al.

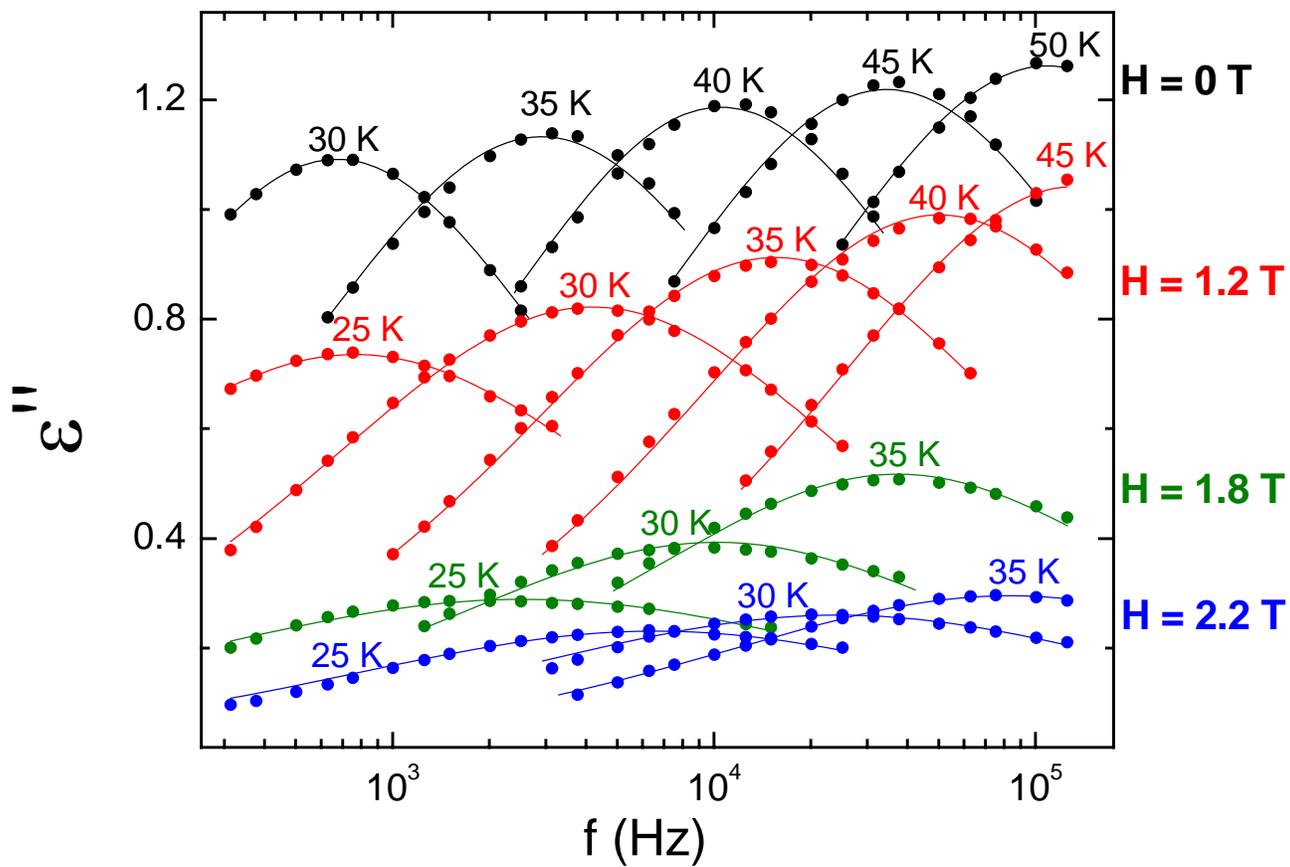

Fig.4 Freitas et al.

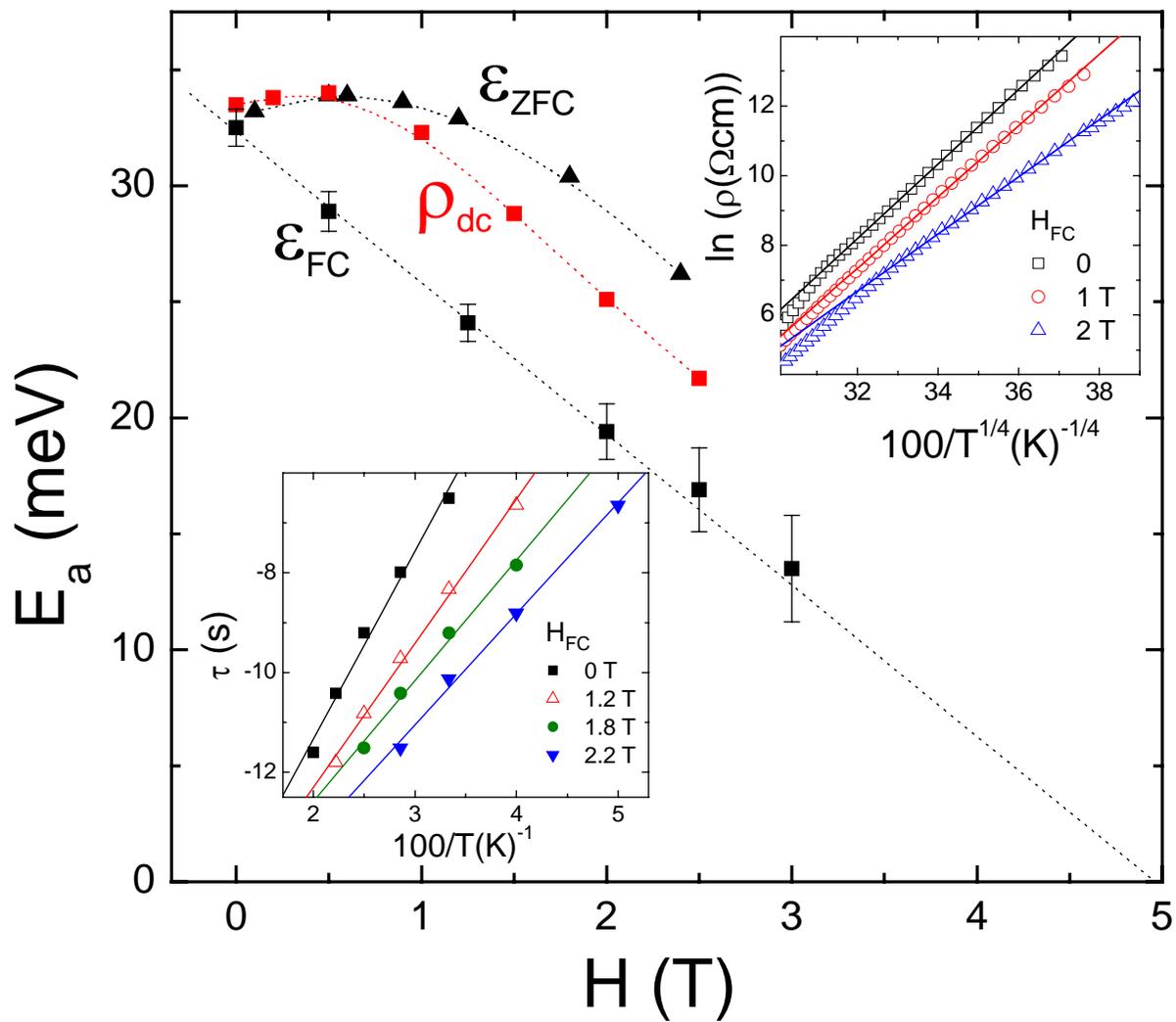

Fig.5 Freitas et al.

Fig.6 Freitas et al.

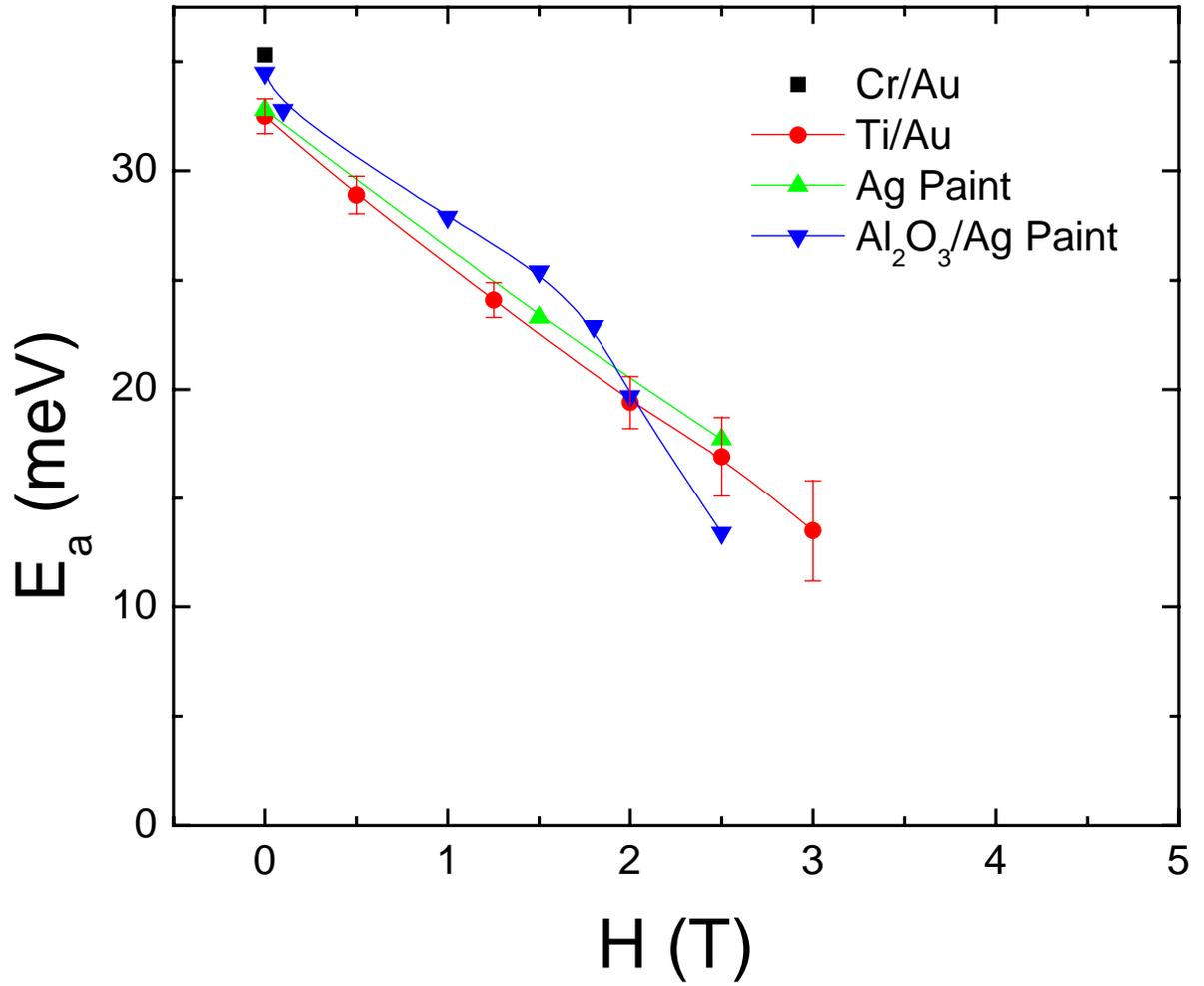